\documentclass[fleqn,10pt]{olplainarticle}
\usepackage[hyphens]{url}
\usepackage{hyperref}
\usepackage{xcolor}
\usepackage{setspace}

\title{Differential fuzz testing to detect tampering in sensor systems and its application to arms control authentication}

\author[1,a,b]{Jayson R. Vavrek}
\author[2,a]{Luozhong Zhou}
\author[3]{Joshua Boverhof}
\author[2]{Elisa R. Heymann}
\author[2]{Barton P. Miller}
\author[3,4]{Sean Peisert}

\affil[1]{Nuclear Science Division, Lawrence Berkeley National Laboratory, Berkeley, CA, 94720, USA}
\affil[2]{Computer Sciences Department, University of Wisconsin-Madison, Madison, WI, 53706, USA}
\affil[3]{Computing Sciences Research, Lawrence Berkeley National Laboratory, Berkeley, CA, 94720, USA}
\affil[4]{Computer Science and Public Health Departments, University of California, Davis, Davis, CA, 95616, USA}
\affil[a]{These authors contributed equally to this work}
\affil[b]{Corresponding author: \href{mailto:jvavrek@lbl.gov}{jvavrek@lbl.gov}}


\begin{abstract}
In future nuclear arms control treaties, it will be necessary to authenticate the hardware and software components of verification measurement systems, i.e., to ensure these systems are functioning as intended and have not been tampered with by malicious actors.
While methods such as source code hashing and static analysis can help verify the integrity of software components, they may not be capable of detecting tampering with environment variables, external libraries, or the firmware and hardware of radiation measurement systems.
In this article, we introduce the concept of physical differential fuzz testing as a challenge-response-style tamper indicator that can holistically and simultaneously test all the above components in a cyber-physical system.
In essence, we randomly sample (or ``fuzz'') the untampered system's parameter space, including both normal and off-normal parameter values, and consider the time series of outputs as the baseline signature of the system.
Re-running the same input sequence on a untampered system will produce an output sequence consistent with this baseline, while running the same input sequence on a tampered system will produce a modified output sequence and raise an alarm.
We then apply this concept to authenticating the radiation measurement equipment in nuclear weapon verification systems and conduct demonstration fuzz testing measurements with a sodium iodide (NaI) gamma ray spectrometer.
Because there is Poisson noise in the measured output spectra, we also use a mechanism for comparing inherently noisy or stochastic fuzzing output sequences.
We show that physical differential fuzz testing can detect two types of tamper attempts, and conclude that it is a promising framework for authenticating future cyber-physical systems in nuclear arms control, safeguards, and beyond.

\end{abstract}

\begin{document}
\flushbottom
\maketitle
\thispagestyle{empty}

\doublespacing

\section{Introduction}

\subsection{Nuclear arms control}
Nuclear arms control treaties attempt to place limits on nations' nuclear arsenals, typically in terms of the number of deployed nuclear weapons.
For instance, the New START Treaty (the only nuclear arms control agreement currently in force) limits the United States and Russia to 1550 deployed nuclear warheads each~\cite{newstart_nti}.
Future nuclear arms control treaties are likely to require some mechanism for directly verifying nuclear warheads, and several groups in the past decade have proposed methods for verifying the authenticity of nuclear warheads without revealing sensitive information~\cite{glaser2014zero, philippe2016physical, kemp2016physical, vavrek2018experimental, hecla2018nuclear, engel2019physically, marleau2017investigation}.
Because there is a strong incentive for the weapon owner (often called the \textit{host} or \textit{monitored party}) to cheat the verification measurement and secretly retain their nuclear weapons or material, these proposed methods must be able to detect hoax objects and must be resistant to tampering by a malicious host.
In this latter tampering problem, the term \textit{authentication} is given to the process(es) by which the inspector (or \textit{monitoring party}) gains confidence that the verification system functions as expected and has not been tampered with.
The inspector may be an independent third party such as the International Atomic Energy Agency (IAEA), a third-party country not party to the treaty, or, as in the case of New START, another signatory of the treaty.
We note that the use of the term \textit{authentication} here is distinct from its usual computer security sense of validating a user's identity.
By contrast, \textit{certification} is the process by which the host gains confidence in the verification system; in particular, that the system (a) does not present a safety hazard around nuclear weapons or other sensitive infrastructure, and (b) that it does not {allow the inspector to store or transmit any sensitive information.
Because certification may involve intrusive access to the verification system, perhaps even without the inspector present, it is a prime opportunity for tampering by the host.
Under New START, for instance, if the verification system is inspector-provided, then the inspector must send (copies of) the verification equipment to the host country for certification $30$~days in advance of an inspection, the host may partially disassemble the to-be-used equipment in an eight-hour examination (albeit with the inspector present), and the equipment is stored in a host-controlled facility between uses~\cite[p.~43--44]{newstart_inspection_annex}.

Moreover, under New START, the authentication process is rather limited, largely due to the fact that the verification measurements themselves are also rather limited by the aforementioned concerns over information security.
Specifically, only simple neutron count rate verification measurements with He-3 neutron detectors paired with commercial counters are permitted, and even these are primarily for confirming the \textit{absence} of nuclear material on certain delivery systems~\cite[p.~36]{newstart_inspection_annex}.
In turn, the New START authentication procedure is primarily limited to \textit{functional testing}, i.e., running basic checks that the verification system's neutron detectors can indeed detect elevated neutron count rates from calibration sources, with little to no anti-tamper testing of the counter's underlying micro-controller software.

If direct warhead verification measurements are to be made as part of a future treaty, more stringent authentication procedures will be required, especially since the scale and complexity of the verification system will grow beyond New START levels~\cite{brotz2019trusting, hamel2018next}.
As such, recent trends in authentication have focused on simple (and joint) hardware design (e.g.,~\cite{kutt2019vintage}) and clear software design choices~\cite{white2001increasing}.
In the latter case, recommendations include using C rather than C++~\cite{white2005computer}, rigorously formatting and linting source code~\cite{weber2023equipment}, and applying formal methods~\cite{evans2015software}.
In addition, several field demonstrations have been carried out, including the 2022 Data Authentication Demonstration~\cite{brotz2023data} and (though it focused more on verification than authentication) the 2020 LETTERPRESS exercise~\cite{quad2020letterpress}.
Despite these efforts and much other work~\cite{tolk1997verification, kouzes2001authentication, hauck2010defining, macarthur2012simultaneous, white2008tools, white2015trends}, authentication remains an open problem in arms control.

\subsection{Fuzz random testing}

Fuzz testing is a random testing technique that was first used in the late 1980's as a simplified way of finding bugs in software programs~\cite{miller1990empirical, klees2018evaluating, miller2020relevance}.
Fuzz testing explores the state space of a program by probing it with random inputs called ``fuzz'' and observing the results, checking for unexpected results that suggest the presence of underlying bugs.
In its original form, it was
\begin{itemize}
    \item \emph{black box}: used no information about the structure of the program being tested;
    \item \emph{unstructured}: used simply random strings without any predetermined syntax or format; and
    \item \emph{generational}: created each new random input for each test run, as opposed to \emph{mutational} that would randomly modify previous test inputs.
\end{itemize} 
The key distinguishing characteristic of fuzz testing is its \emph{oracle}, the mechanism for automatically checking if the program being tested produced the correct answer when given a test input.
Under previous testing methods, the automated construction of a test oracle can be quite challenging, and can rely on assumptions about correct system behavior.
Fuzz testing reduces the oracle to a simple criterion: did the program crash or did it exit normally?
Such a simple criterion allows fuzz testing to be easily and rapidly applied to a wide variety of programs.
When a program crashes under fuzz testing, it is a sign that the random input drove the program into an unexpected internal state.
Fuzz testing does not replace thorough \emph{functional testing}~\cite{kaner1999testing}, which utilizes inputs specified by software test cases, but it is good at exposing many coding errors, and does not depend on assumptions about correct system behavior.

\emph{Differential testing}~\cite{mckeeman1998differential} is a technique where two implementations of a program specification are run and the outputs are compared to identify differences in behavior.
These differences will typically indicate a bug in one of the implementations.
In this case, a test generator with knowledge of the systems and specification can automatically create a set of inputs and the oracle is a simple comparison of the outputs of the two implementations.
\emph{Differential fuzz testing}~\cite{yang2011finding} is a variant of differential testing where the inputs are generated randomly, further simplifying the testing process.

Our use of differential fuzz testing is innovative in that the system inputs are passed to a physical non-deterministic system, resulting in the need to use statistical comparison techniques for the test oracle to deal with noisy outputs.
In addition, this appears to be the first use of differential fuzz testing to detect tampering in a cyber-physical system.

\subsection{Overview and summary of results}

This paper proposes differential fuzz testing as a tamper indicator for cyber-physical systems, with a particular application in authenticating radiation measurement equipment for arms control treaties.
The high-level goal is to randomly explore the input parameter space of a reference cyber-physical system (i.e., one known not to have been tampered with), recording the output of the system.
This output becomes the baseline or oracle signature.
Later, when evaluating the system for tampering, the same inputs are replayed and outputs compared to the baseline.
Sufficiently different outputs indicate that system has been tampered with.
In this way, we apply differential fuzz testing to the same system at two different points in time, rather than two separate implementations as in traditional differential fuzz testing.
Moreover, physical differential fuzz testing can be viewed as another realization of the randomized challenge/response format that has proven useful in related nuclear verification concepts, e.g., Refs.~\cite{kemp2016physical, tobisch2023remote}.
A particular advantage of our technique is that it can test the entire cyber-physical system---source code, environment variables, libraries, firmware, and hardware.
A simple hash on an analysis source code file will, for instance, fail to detect tampering of an environment variable that specifies which analysis library to load.

Our main contributions are twofold: first, the application of fuzz testing to generate a signature of a system, and by extension, the use of differential fuzz testing as a tamper indicator; and second, the application of differential fuzz testing to a physical, non-deterministic system such as a radiation detector.
While our focus is on authenticating an arms control verification system, the physical differential fuzz testing concept could equally well be applied to international nuclear safeguards or any other cyber-physical domain.

\section{Methods and Materials}

\subsection{Concept of operations (CONOPS)}\label{sec:conops}

The concept of operations (CONOPS) for any nuclear arms control treaty will be highly dependent on the specific verification/certification/authentication measurements to be made, and will involve careful negotiation over precise technical details.
For the purposes of demonstrating the fuzz testing approach, however, we adopt the following simplified, general CONOPS as a model based on the existing New START authentication/verification system:
\begin{enumerate}
    \item The inspector performs their initial system tests, establishing a baseline ``fuzz sequence'' by randomly sampling the verification system's state space.
    \item The verification system is turned over to the host for certification; tampering may or may not occur.
    \item The inspector and the host jointly make treaty verification measurements.
    \item The host returns the verification equipment to the inspector.
    \item The inspector conducts a second fuzz sequence with the same random inputs as the baseline and compares the two output sequences.
\end{enumerate}
We note that the second fuzz sequence (step~5) could instead (or additionally) be run \textit{before} the treaty verification measurements of step~3 (if permitted by the treaty) in order to guard against attacks designed to delete their own code after tampering with the verification stage.
However, this option may introduce a \emph{recertification} problem, whereby the host must ensure that the inspector cannot modify the system during the second fuzz testing procedure to achieve some malicious goal of their own.
Similarly, it is important that the equipment is returned to the inspector in such a way that does not give the host an opportunity to undo their tampering; this could be achieved for example by having step~4 occur immediately after step~3.
Regardless of the exact details, the key is that the second fuzz sequence occurs after the host relinquishes control over the inspector's equipment.

A similar general CONOPS can also extend our fuzz testing approach to nuclear security/safeguards rather than arms control scenarios, though this is not our focus in the present work.
For instance, in safeguards for non-nuclear-weapons states, inspection equipment is often left behind in locked cabinets at a nuclear power plant or nuclear fuel production facility, and equipment re-authentication currently depends on containment and surveillance methods (e.g., tags, seals, and video cameras).
Fuzz testing could increase confidence that the containment and surveillance has not been defeated, or at least act as an additional deterrent against such tampering.
In nuclear weapons states, nuclear security measures such as portal monitors in power plants or weapons facilities are also routinely employed~\cite{chambers1974portal, swift2020overview}.
Confidence in the integrity of these systems could again be increased by fuzz testing.

More formally, the inspector supplies some sequence of (pseudo-)random ``fuzz'' inputs to establish a signature of the entire measurement setup (including the radiation detector, its firmware, readout software, analysis software, libraries, and environmental variables).
This sequence should be kept secret from the host.
Let $I_i$ denote the $i\textsuperscript{th}$ set of fuzz input parameters, which could include for instance the detector calibration parameters, the system time, etc.
Let $P_i \equiv \mathcal{M}(I_i)$ be the corresponding set of outputs after the baseline or initial measurement process $\mathcal{M}$, and $P'_i \equiv \mathcal{M}'(I_i)$ be the set of outputs after the test measurement process~$\mathcal{M}'$.
Note that, as discussed further in Section~\ref{sec:tamper_detection}, $\mathcal{M}$ and $\mathcal{M}'$ include the stochastic nature of radiation detection.
The $P_i$ and $P'_i$ could include the energy spectra measured by the detector, any statistics computed from the spectra, and perhaps auxiliary metrics such as disk/network/CPU usage during the measurement.
Finally, let $\mathcal{P} \equiv \{ P_1, P_2, \ldots, P_N \}$ (and analogously, $\mathcal{P}'$) denote the sequence of all outputs after making $N$ measurements with fuzz input parameter sets $I_1, I_2, \ldots, I_N$.
To determine whether the system has been tampered with, we require a method for testing whether the measurement processes $\mathcal{M}$ and $\mathcal{M}'$ differ, i.e., whether the output sequences $\mathcal{P}$ and $\mathcal{P}'$ differ.

\subsection{Tamper detection}\label{sec:tamper_detection}
In traditional differential fuzz testing, the system being tested is a deterministic computer program, and thus it would suffice to check whether $P_i =  P'_i$ for all outputs $i = 1,\, \ldots ,\, N$.
If $P_i \neq P'_i$ for any $i$, then one concludes that the system has been tampered with.
However, in our measurement process $\mathcal{M}$, we make measurements of a radioactive source, which is an inherently stochastic process (both in the radioactive decay and in the detection steps).
Thus, even in the absence of tampering, we expect $P_i \neq P'_i$ in general.
Instead of checking whether the sequences $\mathcal{P}$ and $\mathcal{P}'$ are \textit{equal}, then, we need to check whether they are in some sense \textit{statistically consistent} with each other.

Each $P_i$ or $P'_i$ can contain potentially arbitrary output metrics, some of which will be stochastic (e.g., the output gamma ray spectra) and some of which may be deterministic.
In this work, we restrict ourselves to analyzing only the output gamma ray spectra, though we note again that the outputs could also include other useful metrics such as resource usage.
As such, we essentially comparing two time series of histograms, each of which was generated from the same sequence of fuzzed input parameters.
The two time series should match within Poisson statistics, holding everything else constant.
Malicious code or other tampering should therefore manifest via an anomaly in the time series data.

Let $p_i$ and $p'_i$ denote the gamma energy spectra from the outputs $P_i$ and $P'_i$, respectively.
These energy spectra are just histograms with bin counts $p_{ij}$ and $p'_{ij}$, with $j = 1,\, \ldots,\, J=1024$ the number of channels from the detector readout's ADC.
Many different histogram comparison metrics exist, but we adopt a simple modified reduced $\chi^2$ metric:
\begin{align}\label{eq:red_chi2}
    (\chi^2 / \nu)_i \equiv
    \frac{1}{J} \sum_{j=1}^J \frac{(p_{ij} - p'_{ij})^2}{p_{ij} + p'_{ij}}
\end{align}
where we have modified the denominator to include the expected variance in both measurements~\cite{porter2008testing} and use the convention $0 / 0 \equiv 0$ if $p_{ij} = p'_{ij} = 0$.
Since Eq.~\ref{eq:red_chi2} compares the bin count difference to its expected error, if $\chi^2 / \nu \approx 1$, then the histograms $p_i$ and $p'_i$ are said to be statistically consistent.
If $\chi^2 / \nu \ll 1$ , then $p_i$ and $p'_i$ are much more similar than expected given the measurement statistics.
Conversely if $\chi^2 / \nu \gg 1$, then $p_i$ and $p'_i$ differ substantially and are statistically inconsistent.

While Eq.~\ref{eq:red_chi2} compares a single histogram pair $(p_i, p'_i)$, some method is required for determining whether to alarm based on the collection of all~$N$ histogram comparisons.
For the purposes of the demonstrations in Section~\ref{sec:results}, we will simply set fixed~$\chi^2 / \nu$ thresholds and alarm if any histogram comparison exceeds the threshold.
In general, these thresholds will depend on the magnitude of change in the reduced~$\chi^2$ the inspector expects or wishes to be sensitive to, and setting them inherently involves balancing the risk of a false alarm against a false negative.
Thus in a real scenario, the values of the thresholds should be carefully considered.
In addition, as the number of measurements~$N$ becomes large, the likelihood of having at least one false positive naturally increases; a future implementation should also correct for this multiple comparison effect.
Finally, we note that more advanced anomaly detection methods than the reduced~$\chi^2$ metric could be employed, perhaps leveraging machine learning methods, but that the simpler reduced~$\chi^2$ may be easier to negotiate into a treaty.

\subsection{Demonstration radiation detection system}
To demonstrate our fuzz testing authentication approach with real data, we constructed a demonstration radiation detection system using a $4" \times 4" \times 4"$ sodium iodide (NaI) gamma ray detector coupled to a photomultiplier tube (PMT) (see Fig.~\ref{fig:setup_photo}).
The detector is read out via an ORTEC digiBASE to an Intel NUC running Ubuntu Linux, and is partially enclosed by lead bricks to help isolate it from the radiation environment of the surrounding laboratory.
This setup produces timestamped listmode gamma ray data with energies digitized into $1024$ uncalibrated channels, and a (typically quadratic) calibration function can be used to convert the uncalibrated channel numbers to photon energy depositions in keV if necessary.
\begin{figure}[!htbp]
    \centering
    \includegraphics[width=0.9\columnwidth]{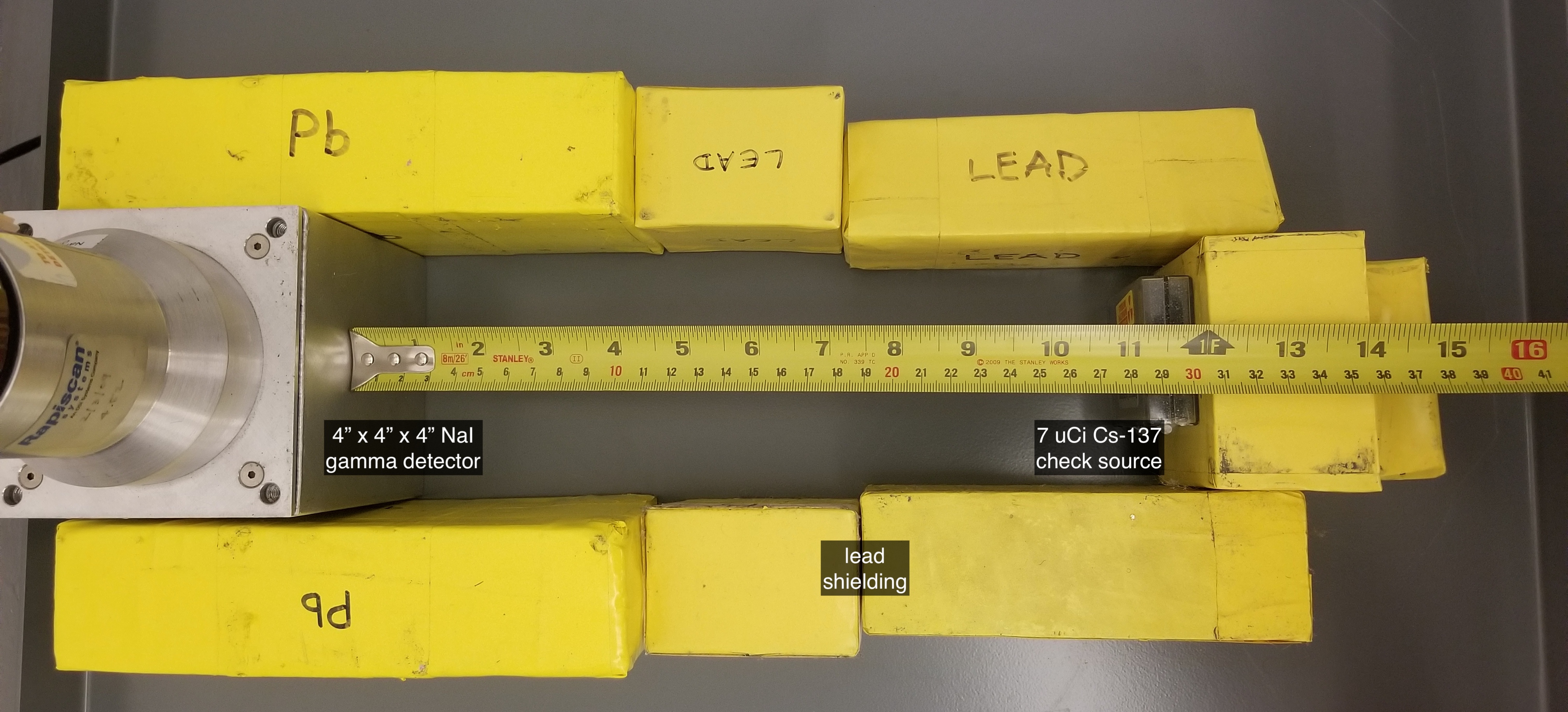}
    \caption{Photograph of the sodium iodide (NaI) detector and source in the demonstration setup.}
    \label{fig:setup_photo}
\end{figure}
We note that this detector configuration is similar to the NaI detector of the Trusted Radiation Identification System (TRIS), which has historically been used to measure the gamma ray spectra of nuclear weapons~\cite{seager2001trusted}.
However, as discussed later in Section~\ref{sec:discussion}, a general-purpose computer such as the Intel NUC used in this proof-of-concept is unlikely to be deployed in a real treaty verification scenario.

A radiation measurement then consists of placing a radiation source near the detector, setting configuration parameters for a pre-set dwell time, starting/stopping acquisition, and writing data to disk.
A list of configuration parameters is given in Table~\ref{tab:config_params}.
In this work, we use a ${\sim}7$~{\textmu}Ci Cs-137 source approximately $30$~cm from the detector face for both the fuzz testing measurements and the proxy treaty verification measurements.
As discussed in Section~\ref{sec:discussion}, in future implementations this source-to-detector distance could also be fuzzed to provide an additional layer of security.
Analysis and attack codes are written in Python3 and make use of the open-source {\tt numpy}~\cite{harris2020array} library.

\begin{table}[!htbp]
    \centering
    \begin{tabular}{l|l|l}
        \textbf{parameter} & \textbf{range} & \textbf{description} \\\hline
        dwell time & $> 0$~s & measurement duration (real time) \\
        high voltage & $100$--$1200$~V & PMT bias voltage\\
        pulse width & $0.75$--$2.0$~{\textmu}s &  detector pulse width in microseconds \\
        fine gain & $0.5$--$1.2\times$ & detector fine gain
    \end{tabular}
    \caption{Select detector parameters available for fuzzing.}
    \label{tab:config_params}
\end{table}

\subsection{Attack pathways}

The space of possible attacks---the \textit{attack surface}---for an arms control verification system is large (especially if software is involved), system-dependent, and difficult to enumerate.
A \textit{useful} attack will achieve some end for the host---such as fooling the system into reporting that a hoax object is a genuine nuclear warhead---without being detected.
The overall threat is based on the fact that the host will have the means, motive, and opportunity to conduct an attack: potentially the resources of an entire nation state, the incentive to create a strategically-useful clandestine stockpile of fissile material, and physical access to the verification system without the inspector present.
Thus, we assume that the host will be able to defeat most existing anti-tamper mechanisms (such as seals, locks, and passwords) without detection and thus will have full access to both the software and hardware of the verification system.
The host will therefore have a wealth of possible mechanisms for attack to choose from, including modifying the analysis/acquisition binary code, third-party libraries, environmental variables, the operating system itself, and any firmware and hardware.

In this work, we consider two relatively simple attack mechanisms, both based on the same point on the attack surface: an attack executed on the listmode gamma ray data stream by overwriting core functions of a loaded {\tt numpy} Python library that will be called later at runtime.
This open-source library is a general-purpose package for manipulating multidimensional arrays, and widely used for scientific computing.
The radiation data acquisition code uses routines in {\tt numpy} to perform array copy ({\tt numpy.copy}) and disk output operations ({\tt numpy.save}). 
To substitute these routines with the attacker's implementation, we create and hide a wrapper script that will be implicitly called with the Python interpreter and update the original function behavior at runtime.
This technique is known as \textit{monkey patching} and is similar to the {\tt Stuxnet} computer worm attack, which replaced a dynamic library (DLL) used in Siemens Step 7 software for controlling centrifuges with a malicious copy~\cite{langner2013kill, zetter2014stuxnet}.
Our attack on the instrument data stream can be executed without modifying the analysis or data acquisition code itself, which would be easily detectable by source code hashing.
In particular, the attacks drop or duplicate listmode gamma ray event data, depending on the attacker's goal.
For example, the attacker may submit an item with some of its gamma-emitting material removed, and duplicate listmode counts to compensate; in this work, we simulate such a reduction of nuclear material by simply increasing the source-to-detector distance.
More sophisticated attacks are discussed in Section~\ref{sec:discussion}.

\subsubsection{Time-based attack}
The first attack is a \textit{time-based attack} in which the host knows the time period in which a verification measurement will take place, implements malicious code to duplicate a random selection of (say) $10 \%$ of gamma ray detection events if the system time is within the known window, and then submits for verification a weapon with $10\%$ of its gamma-ray emitting fissile material removed.
Repeating this process across many weapons, the host could build up a secret reserve of weapons-grade fissile material while appearing to fulfill its treaty obligations.

Fig.~\ref{fig:time-detect} shows an example scenario in which the inspector and host have agreed to make warhead verification measurements on a given Friday between $2$--$3$~pm.
Before sending the inspection system to the host country, the inspector runs an initial fuzz sequence to establish the baseline behavior of the system; one of the fuzzed parameters is the system time.
During certification at the host facility, the host monkey patches the analysis library in order to randomly duplicate listmode gamma ray events with a $10\%$ probability.
To avoid detection by simple source code hashing, the host ensures that this malicious injected script is well-hidden on the file system and that the redirected malicious routine(s) will modify the data only if the system time is between $2$--$3$~pm on Friday.
The host then prepares a hoax weapon with some of its nuclear material removed, such that its gamma emission rate is reduced by $10\%$, and retains the removed nuclear material in secret.
Then, during the designated Friday $2$--$3$~pm window, the inspector and host make treaty verification measurements.
The attack triggers, and the inspector is fooled into believing the correct amount of weapons material has been accounted for.
However, after the treaty verification measurements, the inspector runs another fuzz sequence with the same input parameters as the baseline.
Because the fuzzed parameters include the system time, the malicious code is occasionally re-executed \textit{without} the corresponding reduction in source material, producing an observable change in the output gamma spectra that is detected by the $\chi^2 / \nu$ metric.

\begin{figure}[!htbp]
    \centering
    \includegraphics[width=1.0\columnwidth]{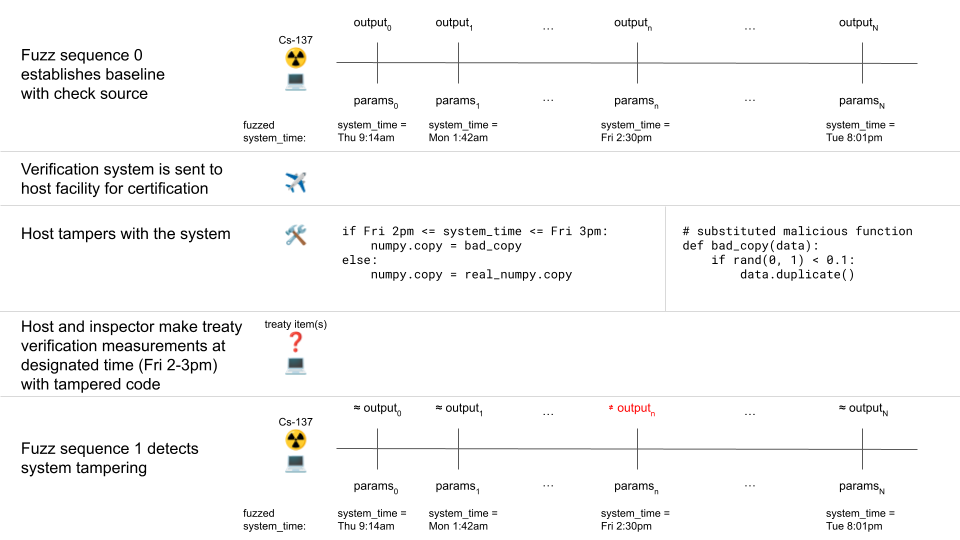}
    \caption{
        Overview of time-based attack detection by fuzzing the system time.
        Here and in Fig.~\ref{fig:count-detect}, the pseudocode is merely representative of the full attack code.
    }
    \label{fig:time-detect}
\end{figure}

\subsubsection{Counts-based attack}
The second attack is a \textit{counts-based attack} in which the attack analyzes the incoming gamma ray data and decides whether and how to execute.
In this example---see Fig.~\ref{fig:count-detect}---the malicious code calculates the total counts in the raw spectrum for each sample and compares it to a predetermined expected total counts.
Any difference in counts could be innocuous and statistical in nature, or could be a result of the attacker having removed (or added) some amount of fissile material.
In the counts-based attack, the attack code identifies the malicious hoax object by a pre-determined off-normal gross counts value, then drops or duplicates listmode count events via the modified {\tt numpy.save} routine while saving to disk to compensate for the difference.
Similar to the time-based attack scenario, the host can use this attack to build a secret reserve of weapons-grade fissile material.

\begin{figure}[!htbp]
    \centering
    \includegraphics[width=1.0\columnwidth]{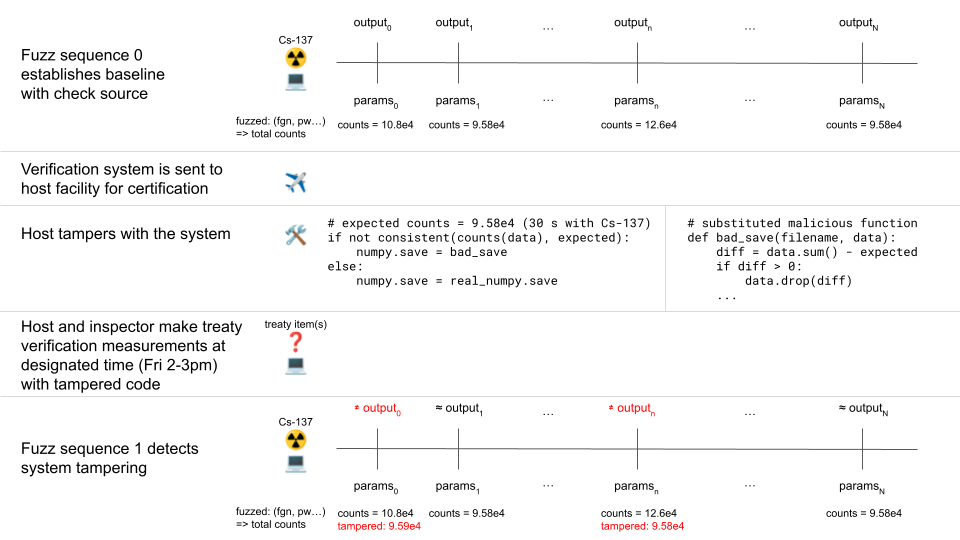}
    \caption{Overview of counts-based attack detection by fuzzing detector parameters that affect total counts.}
    \label{fig:count-detect}
\end{figure}

Once again, this attack can be detected by the fuzz testing procedure.
Fuzzing detector parameters such as the high voltage, fine gain, and pulse width can change the gross counts in the spectrum through various mechanisms such as shifting higher-energy parts of the spectrum outside the dynamic range of the ADC or increasing the number of low-energy counts recorded (regardless of whether the counts are real or are for example electronic noise).
During the second fuzz testing sequence, the malicious logic detects off-normal gross count values due to the off-normal parameter values, rather than the presence of a hoax warhead.
The attack executes, modifying the listmode counts to compensate for the difference in total counts, creating large differences in the $\chi^2 / \nu$~metric between the first and second fuzzing output sequences.

We note for completeness that in our specific attack implementation, for simplicity, we run the attack on every measurement, and not just those deemed ``off-normal''.
Because we \textit{randomly} drop or duplicate listmode counts to make the gross counts more consistent with (but not exactly equal to) the expected number of counts, the modified gross counts still exhibit stochastic fluctuations, rather than remaining constant (which would be a very obvious indication of tampering).
The distribution of fluctuations will however deviate from that expected from Poisson statistics, providing a second potential method to detect the attack (though we do not explore it here).

\section{Demonstration results}\label{sec:results}

\subsection{Time-based attack}

Figs.~\ref{fig:time-chi2} and \ref{fig:time-spectra} show how our fuzz testing procedure detects the time-based attack on our demonstration NaI gamma spectroscopy system.
We first establish the baseline system behavior via $100$ separate radiation measurements with a constant dwell time of $30$~s and PMT high voltage of $1000$~V, while fuzzing the pulse width uniformly (pseudo-)randomly between $0.75$ and $2.0$~{\textmu}s, the fine gain between $0.5$ and $1.2$, and the system time between $1$~pm and $4$~pm.
The left panel of Fig.~\ref{fig:time-chi2} shows the $\chi^2 / \nu$ metric as a function of measurement number when the pseudorandom fuzz random inputs are repeated on a system that has not been tampered with; all $\chi^2 / \nu$ values are below the threshold of~$2$, and no alarm is raised.
Conversely, the right panel shows the system behavior when the fuzz random inputs are repeated on the tampered system.
Approximately one third of the fuzzed system times fall within the $2$--$3$~pm attack window and trigger the malicious code that artificially duplicates counts with a $10\%$ probability.
As shown in Fig.~\ref{fig:time-spectra}, this leads to a corresponding $10\%$ increase in counts across the entire energy spectrum, and a $\chi^2 / \nu > 2$, raising an alarm.
In this simple example, the chosen threshold of $\chi^2 / \nu = 2$ results in a false alarm rate of $0$ and a false negative rate of $1\%$.

\begin{figure}[!htbp]
    \centering
    \includegraphics[width=0.9\columnwidth]{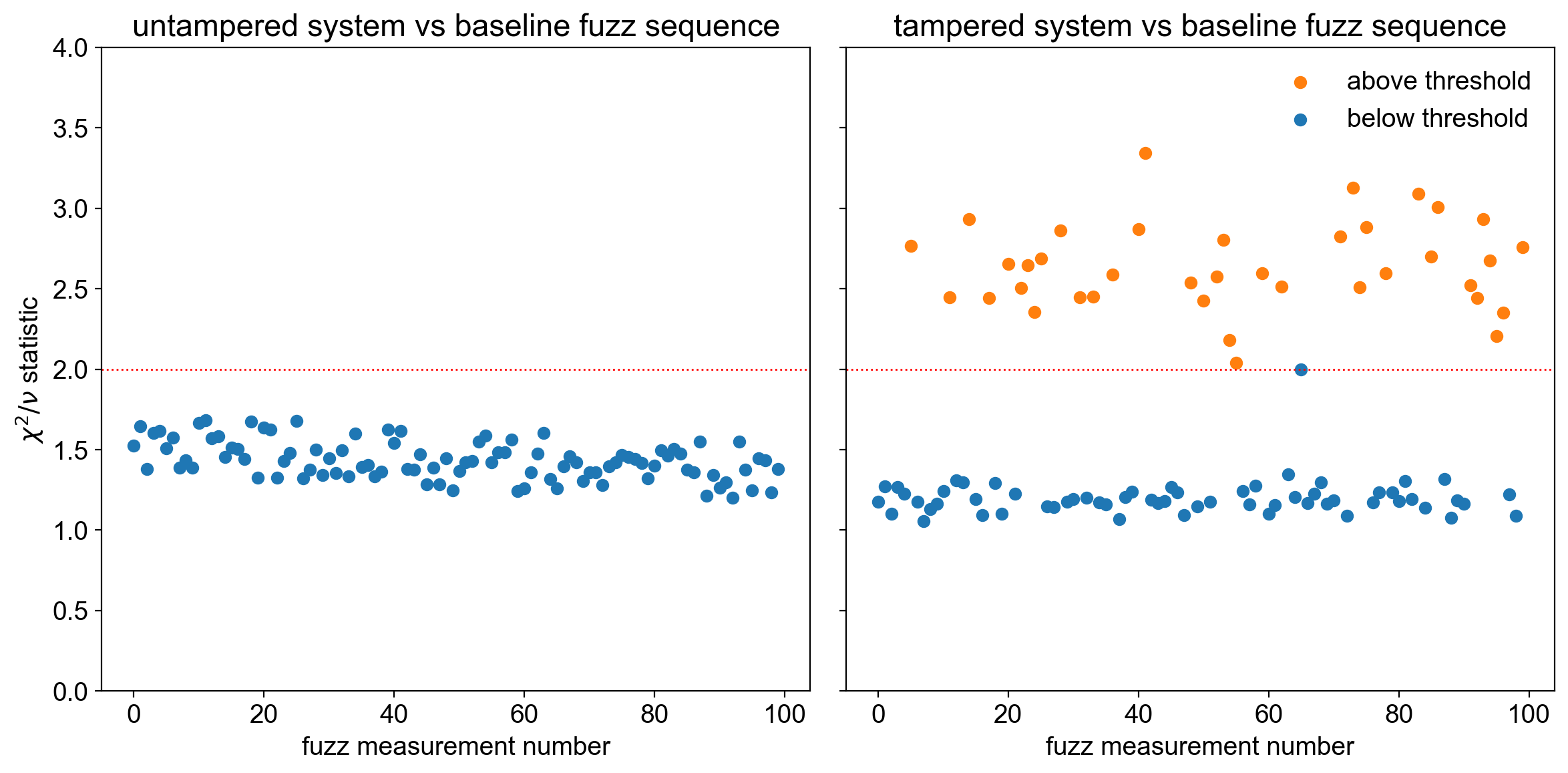}
    \caption{
        Detection of the time-based attack via fuzz testing.
        Left: sequence of $\chi^2 / \nu$ values from an untampered system.
        Right: sequence of $\chi^2 / \nu$ values from a tampered system, with $38$ out of $100$ values crossing the $\chi^2 / \nu = 2$ alarm threshold.
    }
    \label{fig:time-chi2}
\end{figure}

\begin{figure}[!htbp]
    \centering
    \includegraphics[width=0.9\columnwidth]{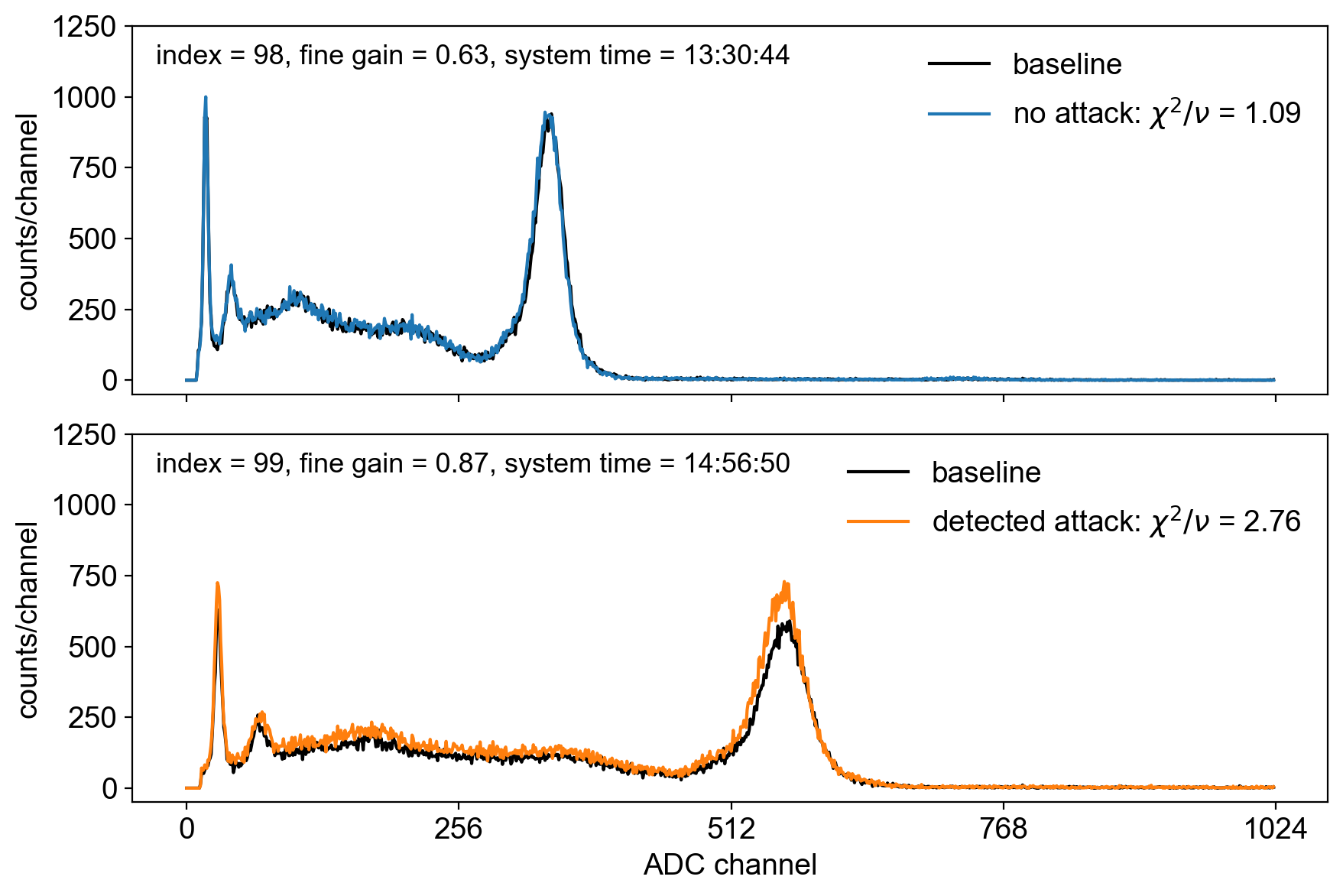}
    \caption{
        Comparison of two fuzz testing measurement pairs from the tampered system (the last two points in the right panel of Fig.~\ref{fig:time-chi2}).
        The tampered system does not execute (does execute) the time-based attack in the top (bottom) panels, respectively, due to the fuzzed system times, leading to observably different $\chi^2 / \nu$ metrics.
    }
    \label{fig:time-spectra}
\end{figure}

\subsection{Counts-based attack}

Figs.~\ref{fig:count-chi2} and \ref{fig:count-spectra} show how our fuzz testing procedure detects the counts-based attack on our demonstration NaI gamma spectroscopy system.
We first establish the baseline system behavior similar to the time-based scenario, but no longer consider the system time as one of the fuzzing parameters.
The left panel of Fig.~\ref{fig:count-chi2} shows the $\chi^2 / \nu$ metric as a function of measurement number when the pseudorandom fuzz random inputs are repeated on a system that has not been tampered with; nearly all $\chi^2 / \nu$ values are below $4$, which we choose as our threshold.
This required threshold of~$4$ is larger than the threshold of~$2$ used in the time-based attack, and is likely due to larger environment-dependent NaI+PMT gain shifts between measurement sequences.
The topic of environmental variations will be discussed further in Section~\ref{sec:discussion}.
We also note that the first comparison in the left panel has an anomalously large $\chi^2 / \nu$ that may be the result of spurious behavior related to detector start-up---i.e., a false positive.
The right panel then shows the system behavior when the fuzz random inputs are repeated on the tampered system.
$70\%$ of the fuzzed samples have total counts that significantly deviate from the expected value (here $9.58 \times 10^3$) and trigger the malicious logic in the routine that artificially duplicates or drops counts to produce tampered data.
As shown in Fig.~\ref{fig:count-spectra}, this leads to a corresponding $56\%$ decrease in counts across the entire energy spectrum of (for example) sample index $97$, and a $\chi^2 / \nu \gg 4$, raising an alarm.

\begin{figure}[!htbp]
    \centering
    \includegraphics[width=1.0\columnwidth]{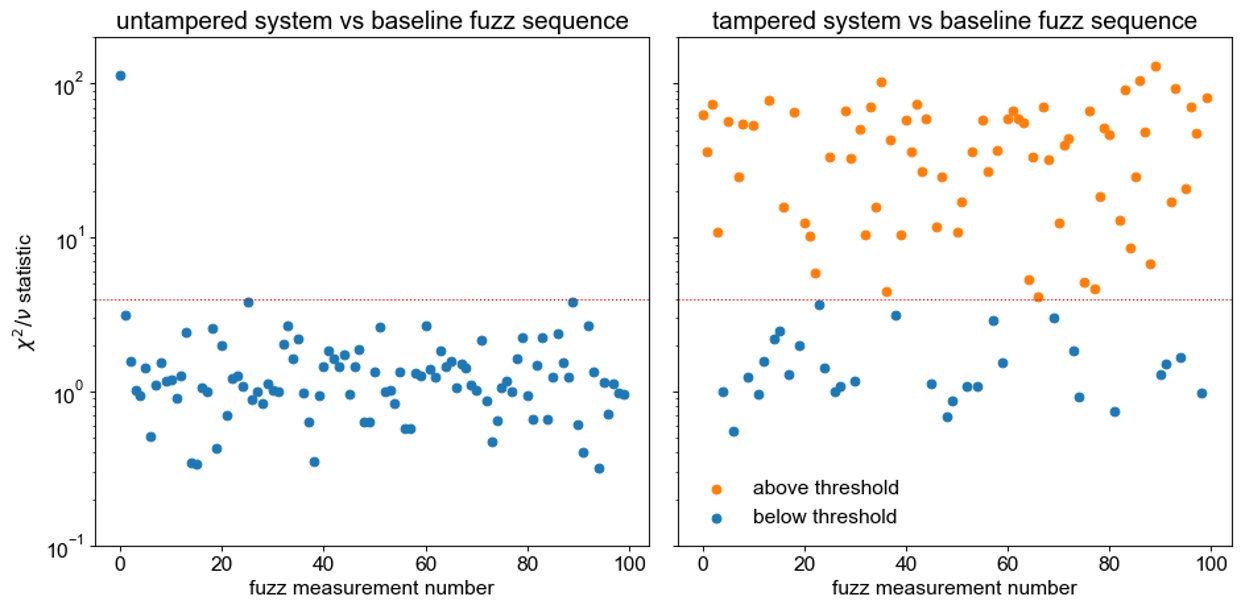}
    \caption{Detection of the counts-based attack via fuzz testing.
        Left: sequence of $\chi^2 / \nu$ values from an untampered system.
        Right: sequence of $\chi^2 / \nu$ values from a tampered system, with $70$ out of $100$ values crossing the $\chi^2 / \nu = 4$ alarm threshold.}
    \label{fig:count-chi2}
\end{figure}

\begin{figure}[!htbp]
    \centering
    \includegraphics[width=0.8\columnwidth]{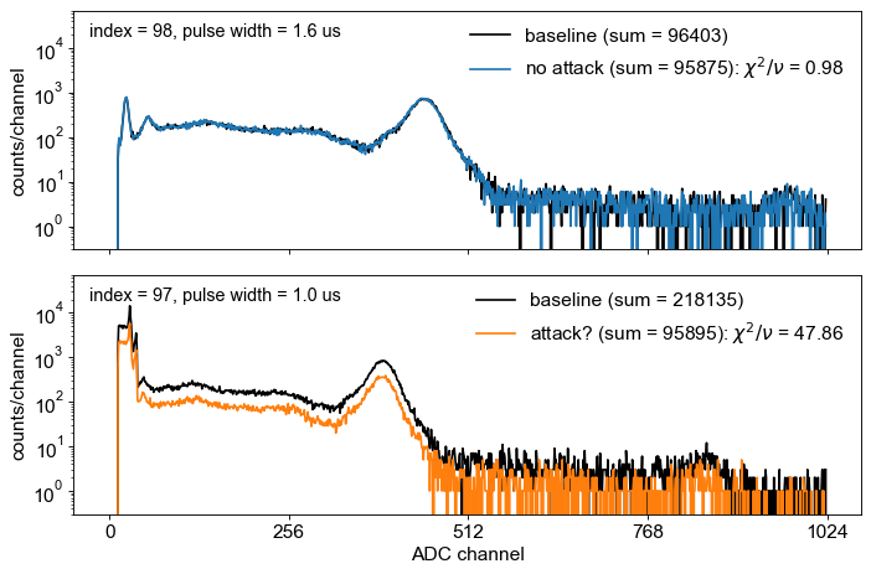}
    \caption{Comparison of two fuzz testing measurement pairs from the tampered system (measurement numbers $97$ and $98$ in the right panel of Fig.~\ref{fig:count-chi2}).
        The tampered system does not execute (does execute) the counts-based attack in the top (bottom) panels, respectively, due to the total counts, leading to observably different $\chi^2 / \nu$ metrics.}
    \label{fig:count-spectra}
\end{figure}

\section{Discussion}\label{sec:discussion}
We have demonstrated the utility of physical differential fuzz testing for detecting certain attacks on arms control verification systems, but we acknowledge some limitations of our current proof-of-concept that should be addressed in future work.

First, more sophisticated attacks may expand the state space of the system and will require more effort to detect.
For instance, the attacker can code their malicious program to only activate under certain specific conditions, also known as a \textit{logic bomb}---historical analogues here include {\tt Stuxnet}~\cite{langner2013kill, zetter2014stuxnet}, the Volkswagen diesel emissions scandal~\cite{schiermeier2015science}, and the arms-control-focused 2015 Underhanded C Contest~\cite{underhanded_c_contest}.
Our time-based attack is a straightforward case in which an activation condition (a given system time range) is present, but is also one of the parameters that is directly fuzzed, and thus the attack is readily detectable.
In our counts-based attack, the attack uses a small amount of ``live'' data (namely the count rate) to determine whether to activate the attack, but is still detected because such conditions are replicated during the fuzz testing.
A more advanced attack might activate only when the system determines via gamma ray spectroscopic analysis that the system is measuring special nuclear material (i.e., weapons-grade U and/or Pu).
Then, if the fuzz testing measurements only use (say) Cs-137 (as they do in this work), the attack will never be triggered during the fuzz testing measurements and will not be detected by the fuzz testing analysis.
It may become necessary, then, to perform the fuzz testing using standardized U/Pu calibration sources, though this will increase the operational challenge compared to a Cs-137 check source.
Similarly, the attack code could be designed to determine whether the system is in fuzz testing mode or weapon verification mode based on the measurement dwell time(s)---the repeated constant $30$~s dwell time used here for simplicity could indicate fuzz testing while longer dwells might indicate weapon verification.
Fuzzing the measurement dwell times, as suggested in Table~\ref{tab:config_params}, may provide an additional layer of ``anti-anti-tamper'' defense.

A malicious host may also attempt a replay attack, in which they construct an \textit{approximate oracle} that provides realistic-looking but synthetic spectra at a wide range of fuzz input parameters, bypassing the detector+readout chain entirely.
Such an attack could be based on a digital twin of the detector system, or may comprise a data-driven interpolation table or other latent space representation.
Both methods will require additional compute resources that may be difficult to hide from the inspector, and will suffer from some level of approximation error that would be interesting to quantify in future work.
Moreover, even a well-crafted and well-hidden replay attack could be discovered by the addition of fuzz input parameters that cannot be controlled by the detection system, and thus cannot be modeled by the attack.
For instance, fuzzing the source-to-detector distance could detect a replay attack that assumes the calibration source is located at a fixed distance of $d=1$~m, since the incident signal intensity varies as $1/d^2$.

Systematic environmental changes between different fuzz sequences present an additional challenge, as in this work we have either (1) assumed that variations in output gamma spectra can only arise from Poisson noise or tampering, or (2) adjusted $\chi^2 / \nu$ thresholds in a somewhat ad-hoc fashion to account for environmental variations.
Our NaI+PMT detection system is particularly sensitive to environmental variations due to the non-linearity of the PMT gain with temperature.
Such variations stretch or squeeze the gamma spectra in channel space, potentially causing spectroscopic misalignments and thus anomalously large $\chi^2 / \nu$ values.
Similarly, variable natural backgrounds across time and measurement locations may also increase spectroscopic disagreements, but the influence of background can be minimized with shielding and strong calibration sources.
In this work, we have tried to reduce the impact of environmental variations by collecting both fuzz sequences in the same location within a few hours, and by setting a large enough alarm threshold such that the environmental variations over this time span do not cross the threshold.
One could also explore gain stabilization methods, but we note that these add complexity to the system and in fact expand the attack surface.
In future arms control treaties, therefore, detector materials with better environmental robustness (such as CdZnTe~\cite{park2010effect}) may be preferred for fuzz testing applications.
More generally, it is important to reduce these kinds of systematic variations in any verification method so as to maintain a strict alarm threshold and thus reduce the opportunity for an attacker to hide an attack within the observed noise.

We emphasize again that our demonstration system is only a proxy for a real arms control verification system, as it is unlikely that a general-purpose personal computer running a full operating system and suite of other applications with a multitude of third-party libraries can pass the authentication and certification requirements for in-field use around nuclear warheads.
While current inspection equipment typically consists of (relatively) simple analogue electronic counters designed for a specific measurement, more advanced verification methods could require (say) a small gamma spectroscopy software library running on a microcontroller or edge computer.
If software is present, the source code, binaries, and compiler should be available to the inspector for (necessary but not sufficient) authentication checks against both straightforward modifications and advanced attacks such as Trusting Trust~\cite{thompson1984reflections} or the {\tt xz} hack~\cite{cox2024xz}.

Although we have focused on comparing the output time series of fuzzed gamma energy spectra, it may also be highly useful to monitor more implicit system signatures during fuzzing.
For instance, comparing the system resource usage and required analysis time across fuzzed sequences could detect more computationally-intensive attacks.
An attack that duplicates listmode counts, for example, may need to use additional compute to globally adjust all timestamps on an ongoing basis so that the photon inter-arrival time distribution is still exponential.
(In contrast, an attack that randomly drops listmode counts does not need to further adjust timestamps, since a random sub-sample of an exponential distribution is still an exponential distribution, albeit with a modified shape parameter.)

It may be that multiple copies of the verification system are produced in order to allow, e.g., the random selection of the system used and/or \textit{post-facto} destructive testing of unused copies by each of the host and inspector as anti-tamper deterrents.
In this case, it may be possible to run different fuzz testing sequences in parallel on the system copies, thereby increasing parameter space coverage for the same amount of measurement time.

Testing may also be accelerated through the use of so-called ``gray-box'' fuzz testing techniques, such as power scheduling and coverage-based feedback design.
However, in the arms control application space especially, we expect that the far-simpler black-box technique (uniformly randomly sampling the fuzz parameter space) will be easiest for treaty partners to agree on.

More generally, further work will be necessary to determine the role of fuzz testing within the overall authentication CONOPS.
In ongoing work, for instance, we intend to integrate our radiation detection system with the Chain of Custody Item Monitor (CoCIM)~\cite{brotz2017non}, a fiber-optic active seal developed for use in warhead verification applications, and deploy fuzz testing on the CoCIM software.
Conversely, the utility of fuzz testing for information barriers (e.g.,~\cite{close2001information, chambers2010uk, yan2015nuclear, kutt2018information, brotz2023development}) may be limited, since they typically reduce sensitive measurement data to $1$--$6$ output bits of non-sensitive information.
We also note that many CONOPS details will likely emphasize ease-of-use for the inspector and not necessarily just the detection power of the fuzz testing approach.
As a simple example, the inspector may not want (or may not be able) to run fuzz testing over millions of parameter combinations over the span of many weeks, thereby limiting the possible coverage of the fuzz testing approach.
In addition, expanding the technique to different systems presents operational challenges.
Although the basic fuzzing concept remains constant across systems, for each new system one must identify a set of fuzzable input parameters that covers a significant portion of the system's state space.
For example, while nearly all radiation detection equipment will have some kind of high voltage and pulse width parameters (Table~\ref{tab:config_params}), warhead verification systems such as TRIS~\cite{seager2001trusted} are typically designed to have few or no \textit{easily-manipulable} user inputs as a security feature, which may make such parameter identification challenging.
Finally, as alluded to in Section~\ref{sec:conops}, fuzz testing does not address, and in some cases may even exacerbate the certification problem, at the expense of addressing the authentication problem.
It is important to emphasize therefore that fuzz testing will not make cheating on an arms control treaty impossible, especially not on its own.
Rather, it should be deployed as one among many tools for making cheating infeasibly difficult and/or expensive.

\section{Conclusions}
We have proposed the use of differential fuzz testing as a tamper indicator of sensor systems, with a specific application to radiation measurement equipment in nuclear arms control verification systems.
Our results show that the fuzz testing method is capable of detecting representative tampering attempts, in particular on the listmode gamma ray data stream.
While a number of refinements remain open for future work, differential fuzz testing of sensor systems is a promising tool for tamper detection in arms control authentication and cyber-physical systems more broadly.

\section*{Acknowledgments}

The authors thank Marco Salathe, Mark Bandstra, John Valentine, and Jay Brotz for useful discussions.

This work was performed under the auspices of the U.S.~Department of Energy by Lawrence Berkeley National Laboratory under Contract DE-AC02-05CH11231.
The project was funded by the U.S.~Department of Energy, National Nuclear Security Administration, Office of Defense Nuclear Nonproliferation Research and Development (DNN R\&D).

This document was prepared as an account of work sponsored by the United States Government.
While this document is believed to contain correct information, neither the United States Government nor any agency thereof, nor the Regents of the University of California, nor any of their employees, makes any warranty, express or implied, or assumes any legal responsibility for the accuracy, completeness, or usefulness of any information, apparatus, product, or process disclosed, or represents that its use would not infringe privately owned rights.
Reference herein to any specific commercial product, process, or service by its trade name, trademark, manufacturer, or otherwise, does not necessarily constitute or imply its endorsement, recommendation, or favoring by the United States Government or any agency thereof, or the Regents of the University of California.
The views and opinions of authors expressed herein do not necessarily state or reflect those of the United States Government or any agency thereof or the Regents of the University of California.

This manuscript has been authored by an author at Lawrence Berkeley National Laboratory under Contract No.~DE-AC02-05CH11231 with the U.S.~Department of Energy.
The U.S.~Government retains, and the publisher, by accepting the article for publication, acknowledges, that the U.S.~Government retains a non-exclusive, paid-up, irrevocable, world-wide license to publish or reproduce the published form of this manuscript, or allow others to do so, for U.S.~Government purposes.

\section*{Data availability}
We intend to make data and sample code available at time of publication.

\section*{Author contributions}

\begin{itemize}
    \item JRV: formal analysis, investigation, methodology, resources, software, visualization, writing---original draft, editing.
    \item LZ: data curation, formal analysis, investigation, software, validation, visualization, writing---original draft, editing.
    \item JB: data curation, software, editing.
    \item EH: supervision, editing.
    \item BPM: co-principal investigator; conceptualization, supervision, writing---original draft, editing.
    \item SP: principal investigator; conceptualization, funding acquisition, supervision, writing---original draft, editing.
\end{itemize}

\section*{Competing interests}
The authors declare no competing interests.

\bibliographystyle{IEEEtran}
\bibliography{biblio}

\end{document}